\documentclass[reprint,amsmath,amssymb,aps,floatfix]{revtex4-2}

\usepackage{graphicx}% Include figure files
\usepackage{dcolumn}% Align table columns on decimal point
\usepackage{bm}% bold math
\usepackage{esvect}
\usepackage{amsmath}
\usepackage{xcolor}
\usepackage[mathscr]{euscript}
\usepackage[italicdiff]{physics}
\usepackage{mathtools}
\usepackage[colorlinks=true,linkcolor=blue,citecolor=blue]{hyperref}% add hypertext capabilities
\def\orcid#1{\kern .08em\href{https://orcid.org/#1}{\includegraphics[keepaspectratio,width=1em]{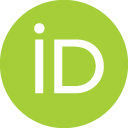}}} 

\begin{document}

\preprint{APS/123-QED}

\title{Information engine with feedback delay based on a two level system}
\author{Kiran V \orcid{0000-0003-3960-2589}}
\email{kiran.vktm@gmail.com, p20180029@goa.bits-pilani.ac.in}
\affiliation{Department of Physics, BITS Pilani K K Birla Goa Campus, 
Zuarinagar 403726, Goa, India}
\author{Toby Joseph \orcid{0000-0001-6682-9223}}
\email{toby@goa.bits-pilani.ac.in}
\affiliation{Department of Physics, BITS Pilani K K Birla Goa Campus, 
Zuarinagar 403726, Goa, India}

\date{\today}

\begin{abstract}

An information engine based on a two level system in contact with a thermal reservoir
is studied analytically. The model incorporates delay time between the 
measurement of the state of the system and the feedback. The engine efficiency and
work extracted per cycle are studied as a function of delay time and energy spacing
between the two levels. It is found that the range of delay time over which one can
extract work from the information engine increases with temperature. For delay times
comparable to the relaxation time, efficiency and work per cycle are both maximum when 
$k_B T \approx 2 U_0$, the energy difference between the levels. Generalized Jarzynski equality 
and the generalized integral fluctuation theorem are explicitly verified for the model. 
The results from the model are compared with the simulation results for a feedback 
engine based on a particle moving in a $1D$ square potential. The variation of efficiency, 
work per cycle and efficacy with the delay time is compared using relaxation time in the 
two state model as the fitting parameter and leads to good fit.

%\begin{description}
%use the optional argument of the \verb+\item+ command to give the category of each %item. 
%\end{description}
\end{abstract}

\maketitle

%\tableofcontents

\section{Introduction}
 
An information engine uses the information gained from a measurement of the system to
extract work from thermal fluctuations \cite{ClerkMaxwell:1871:TH,Maruyama2009}. Historically it 
was Maxwell who first suggested a thought experiment involving a demon whose measurements of molecular velocities
can be used to transfer heat from a cold to a hot body thus violating the second law of
thermodynamics. It is now generally accepted that there is no violation of the second law if
one considers the entropy generation associated with erasure of memory involved in
the measurement process \cite{Landauer1961,bennett1982thermodynamics,Maruyama2009,sagawa2009minimal,leff2003maxwell}. 
For alternative views, see the following references \cite{earman1998exorcist,earman1999exorcist,
hemmo2010maxwell,norton2011waiting,kish2012energy}. Recently information engine or Maxwell's demon, as it
is usually referred to, has been implemented experimentally in a variety of systems both classical
\cite{toyabe2010experimental,berut2012experimental, saha2021maximizing,paneru2018lossless} and quantum 
\cite{koski2014_zilard_experimental,PhysRevResearch.2.032025,averin2011maxwell,
camati2016experimental,naghiloo2018information,chida2017power}.\\

Stochastic thermodynamics deals with study of thermodynamics of small systems where fluctuations
dominate \cite{Jarzynski2013}. Several fluctuation theorems have offered valuable insights into 
the production of entropy and the statistical connections between work and free energy for systems 
operating significantly beyond equilibrium \cite{evans1993probability,gallavotti1995dynamical,
Jarzynski1997,crooks1999entropy}. These results in stochastic thermodynamics has been generalized to the
cases when there is measurement and feedback during the process \cite{Parrondo2015,sagawa2010generalized}. 
For example, the Jarzynski relation which connects the fluctuations in work, $W$, during a non-equilibrium process 
to the free energy difference, $\Delta F$, between the final and initial equilibrium states 
\cite{Jarzynski1997} has been modified to a form, 
\begin{equation}
\displaystyle \left<e^{\beta(\Delta F - W)}\right> = \gamma,
\label{gjr}
\end{equation}
where $\beta$ is the inverse temperature and the angular brackets represent average over multiple trajectories of
the system starting from the equilibrium distribution.
This relation, known as the Generalized Jarzynski Relation (GJR), is valid for non-equilibrium processes 
incorporating feedback mechanism. The right hand side of GJR, $\gamma$ (referred to as efficacy), is the 
sum of the probabilities of observing the time reversed trajectories in the time reversed protocols for 
all possible protocols. $\gamma$ is a measure of the reversibility of the process. The largest value that $\gamma$ 
can take is the number of outcomes in the measurement process and is attained for a fully reversible process.  
In the absence of feedback, $\gamma = 1$ and the GJR reduces to the usual Jarzynski relation \cite{Jarzynski1997}.
The experimental verification of GJR has been done for few systems \cite{toyabe2010experimental,Koski2014,
Paneru2020,paneru2020colloidal}. In the case of processes involving precise measurements (error-free) and 
feedback mechanisms, we can establish a Generalized Integral Fluctuation Theorem (GIFT) expressed as 
$\left<e^{\beta(\Delta F - W) - I + I_u}\right> = 1$. Here, $I$ represents the information acquired 
during the measurement process, and $I_u$ is the unavailable information to be determined through 
the time-reversed process \cite{ashida2014general}.\\

In the context of an information engine, efficiency quantifies the degree of conversion of information to work. 
High efficiency requires a slow process thus compromises on power. Thus it is important to tune the
engine parameters such that efficiency and power are as required. Many recent works 
have investigated methods to enhance the efficiency and power of information engine, both in experiments 
\cite{saha2021maximizing,paneru2018optimal,rico2021dissipation} and in theoretical studies 
\cite{lucero2021maximal,dinis2020extracting,pal2014extracting}.  Information engines based on colloidal 
particle moving through a harmonic potential \cite{abreu2011extracting, bauer2012efficiency, paneru2018lossless} 
as well as periodic potentials \cite{toyabe2010experimental, PhysRevE.106.054146} have been  studied.
These studies look into the possibility of extracting work or converting the information about 
the position of the particle into work with the help of a feedback scheme. An information engine based
on a two level system where the state of the system is measured and feedback is effected has been theoretically 
studied \cite{Um_2015}. These simple information engine systems offer ways to understand the optimisation schemes. \\

In this study, we perform an analytical investigation of a two-state information engine that is in 
contact with a heat bath. The model is similar to the one studied by Jaegon et al \cite{Um_2015} but differs
in that in the current model there is a feedback delay between the measurement and feedback. The analytical
results are derived by assuming that the cycle time of the engine is large compared to the relaxation time of the system. 
The feedback time and the energy difference between the two states 
are the two parameters with respect to which the efficiency and work per cycle of the engine are studied. 
We compare the analytical findings with the numerical results 
obtained from the simulation of a particle moving within a one-dimensional periodic square potential. Over-damped
Langevin dynamics is used to simulate the motion of the particle.
Further, the generalized fluctuation relations of stochastic thermodynamics for this system are verified. \\

% It's worth noting that the 
% simulated square potential is not a perfect 
% representation; deviations may arise due to the inherent limitations in modeling the relaxation process 
% and minor errors associated with modeling the square potential.\\

The paper is structured as follows: The model for the information engine is introduced in the next section 
and the assumptions and parameters of the model are defined. In Sec. {\color{blue}\ref{sec2}}, we start with the study of 
information engine without feedback delay (Sec. {\color{blue}\ref{sec2_1}}) and then generalize to one that 
incorporates feedback delay time (Sec. {\color{blue}\ref{sec2_2}}). The engine performance parameters are worked 
out and the fluctuation theorems are verified.
% We also present the corresponding outcomes. 
% For each model, we define efficiency and work per cycle 
% while also verifying generalized fluctuation relations. Furthermore, we analyze the efficiency and 
% work per cycle concerning the state's energy. In the model featuring feedback delay includes a 
% validation of the GJR for various feedback delay values. 
Sec. {\color{blue}\ref{sec2_3}}  provides a comparison between the analytical results and the results 
obtained from the simulation of the particle moving in periodic square potential. Finally, 
in Sec. {\color{blue}\ref{sec3}}, we offer a summary of our findings and engage in a discussion of the results.

\section{The Model}
The information engine consists of a two level system in contact with
a thermal reservoir at temperature $T$. The energies of the higher energy state
(up state) and the lower energy state (down state) of the system are $U_0$ and $-U_0$ respectively.
Also present as a part of the information engine is an observer (Maxwell's demon)
who measures the state of the system at regular intervals of time, $t = n \alpha$ ($n$ is an integer),
and implements a feedback process depending on the outcome of the measurement.
The feedback process is as follows: If the system is measured to be in the up state
in the $n^{\rm th}$ measurement, the demon flips the state of the system to down state 
at a time $t = n \alpha + \epsilon$, with $\epsilon < \alpha$. $\epsilon$ is the feedback
delay time. If the system is measured to be in the down state, no feedback is initiated. 

%The measurement carried out by the demon is prone to error and we assume an error of the
%form, $P(X = D| M = U) = P(X = U | M = D) = \delta$. That is, the probability that the actual
%state (denoted by $X$) is down given that the measurement outcome (denoted by $M$) was up is
%$\delta$ and vice versa. In general, these two errors need not be the same but we assume it
%for simplicity. It follows that $P(X = D | M = D) = P(X = U | M = U) = 1 - \delta$. 

\begin{figure}
\centering
\includegraphics[scale=0.60]{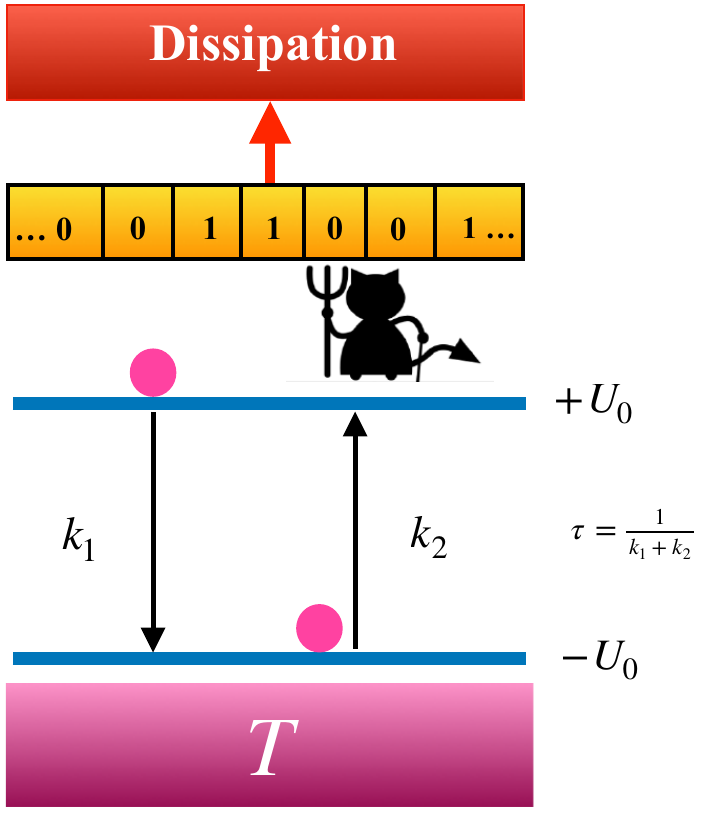}
\caption{The information engine comprises a two-level system in thermal contact with a reservoir 
at temperature $T$. The system's higher energy state (referred to as the 'up state') possesses an energy 
$+U_0$, while the lower energy state (the 'down state') has an energy of $-U_0$. The model consists of a 
device (demon) with a finite memory which measures the system's energy state at a regular interval of 
time and records the measurement outcome in the memory. Depending on the measurement outcome, 
the demon initiates the feedback protocol after a delay time $\epsilon$. The relaxation time $\tau$
of the system is characterized by the transition rates $k_1$ and $k_2$.}
\label{model}
\end{figure}

The master equation for the process is given by
\begin{eqnarray}
    \frac{dp_u(t)}{dt} &=& -k_1 p_u + k_2 p_d \nonumber \\
    \frac{dp_d(t)}{dt} &=&  k_1 p_u - k_2 p_d,
\end{eqnarray}
where $p_u$ and $p_d$ are the probabilities for finding the system in up and down states respectively
and $k_1$ and $k_2$ are the rates of transition between the states (see Fig. \ref{model}). 
Detailed balance condition in equilibrium dictates that $\frac{k_2}{k_1} = e^{-2 \beta U_0}$, 
where $\beta = \frac{1}{k_B T}$. We shall work in energy unit where $k_B T = 1$. 
The relaxation time for the process is, $\tau = \frac{1}{k_1 + k_2}$. 
Note that for the case when the measurement outcome is up state, the master equation has to 
be integrated in two time segments: from $t = n \alpha$ to $t = n \alpha + \epsilon$ and 
then from $t = n \alpha + \epsilon$ to $t = (n + 1) \alpha$. This is because, if the the 
measurement gives up state as the outcome, the state will be flipped after a delay time 
of $\epsilon$.

\section{Results and analysis}\label{sec2}
\label{Results}
In the analysis that follows, we shall assume that the time between the state flip and the
next measurement time, $\alpha - \epsilon$, is much larger than the relaxation time $\tau$. 
This would imply that the system is in equilibrium at the beginning of each cycle. We first
work out the simpler case when $\epsilon = 0$ (immediate feedback with no delay time). 
Subsequently, we relax this constraint and work out the results for the more general case.
\begin{figure}
\includegraphics[scale=0.45]{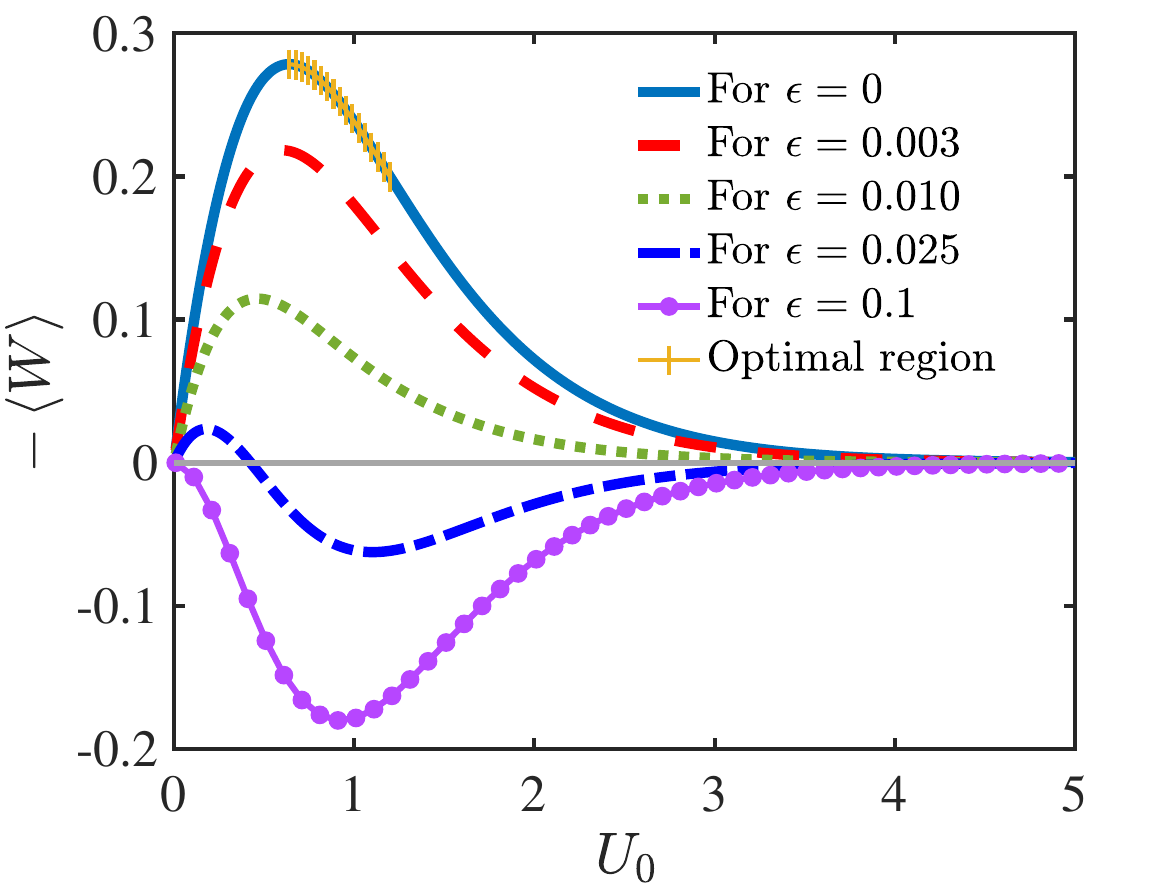}
\caption{\label{fig:U_pow} Variation of WPC for the information engine as a function of $U_0$
(given in units of $k_B T$). Different curves correspond to varying delay times. The value of 
relaxation time for all the curves is, $\tau = 0.02$. The blue solid curve corresponds to the
case of $\epsilon = 0$. The shaded yellow section on the solid blue curve corresponds to range of operation
where WPC and efficiency are both appreciable.}
\end{figure}

\subsection{Feedback engine with no delay time ($\epsilon = 0$)}\label{sec2_1}
We consider here the case when the feedback is implemented right after the measurement.
The probability for spotting the particle in the up state during the measurement is,
\begin{equation}
\displaystyle p_{u}^{eq} = \frac{e^{-\beta U_0}}{2\cosh{(\beta U_0} )},
\label{p_upper}
\end{equation}
which is the equilibrium distribution. Average information gathered during the measurement is,
\begin{equation}
\displaystyle \left<I\right> = -p_{u}^{eq}\ln{p_{u}^{eq}} - (1-p_{u}^{eq})\ln{(1 - p_{u}^{eq})} \;.
\label{information}
\end{equation}
This is related to the cost of running the information engine. Processing this information requires
a minimum of $k_B T \left<I\right>$ of energy, associated with resetting the memory bits
involved in the measurement process. 

\subsubsection{Efficiency and work per cycle}
Since a work of $2 U_0$ is extracted every time the particle is spotted in the up state, 
the average work extracted per cycle is,
\begin{equation}
\displaystyle -\left<W\right> = 2 U_0 p_{u}^{eq}.
\label{work}
\end{equation}
The variation of average work per cycle (WPC) as a function of $U_0$
is shown in Fig. \ref{fig:U_pow} (blue solid curve). The optimal value for $U_0$ at which 
WPC is a maximum is $U_0 = 0.64 \;k_B T$. The fact that $U_0$ has to be of the order of $k_B T$ for optimal
work extraction can be understood as follows: If $U_0$ is much smaller than $k_B T$, the chance
of spotting the system in the up state will be close to $0.5$, but the resultant work extraction 
per flip will be small. For $U_0$ much larger than $k_B T$, the probability of observing the 
system in the up state reduces drastically leading again to low value of WPC. 

\begin{figure}
\includegraphics[scale=0.45]{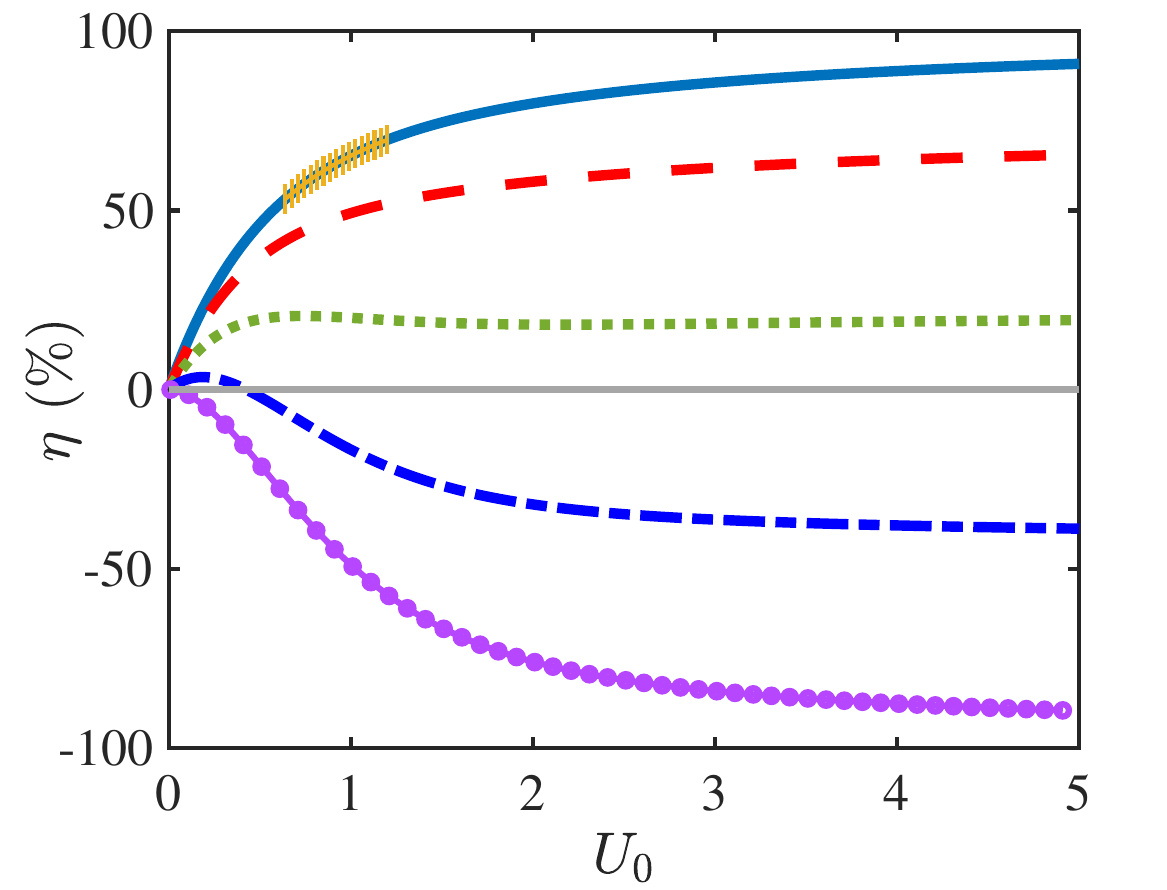}
\caption{\label{fig:U_eta} Variation of efficiency of the information engine as a 
function of $U_0$ (given in units of $k_B T$) for various delay times. 
The legends and value of $\tau$ are the same as used in Fig. \ref{fig:U_pow}.
The shaded yellow section on the solid blue curve corresponds to range of operation
where WPC and efficiency are both appreciable.}
\end{figure}

The efficiency, defined as the ratio of work extracted to the cost of running the engine, is given by
\begin{equation}
\displaystyle \eta = \frac{2 U_0 p_{u}^{eq}}{k_B T \left<I\right>}.
\label{efficiency_1}
\end{equation}
The efficiency is a monotonically increasing function of $U_0$ and saturates for values of
$U_0 \gg k_B T$ as seen in Fig. \ref{fig:U_eta} (blue solid curve). In the limit of 
$U_0 \rightarrow \infty$, $p_{u}^{eq} \sim e^{- \beta U_0}$. 
It is easily seen that in this limit, $\eta \rightarrow 1$. But this maximal
efficiency happens at the expense of WPC going to zero. At
the value of $U_0$ at which WPC is a maximum, $\eta \approx 53 \%$. For optimal choices that do 
not compromise either efficiency or work per cycle, $U_0$ should lie between $0.6$ and $1.2$ 
(the shaded yellow region in the blue solid curve in Figs. \ref{fig:U_pow} and \ref{fig:U_eta}) for the 
case when $\epsilon = 0$.

\subsubsection{Verifying generalized Jarzynski relation}
We compute the right and left hand sides of the GJR (Eq. (\ref{gjr})) separately for verifying the relation. Note
that $\Delta F = 0$ for the system considered because the energy spacing remains the same after the
feedback process. The work variable, $W$, can take two values: (i) $W_1 = -2 U_0$, when particle is observed 
in the up state and (ii) $W_2 = 0$, when the particle is observed in the down state. The corresponding
probabilities are, $P(W_1) = p_{u}^{eq}$ and $P(W_2) = 1 - p_{u}^{eq}$. Thus the left
hand side of the GJR is 
\begin{eqnarray}
\left <e^{-\beta W}\right> &=& e^{2\beta U_0} p_{u}^{eq} + 1 - p_{u}^{eq} \nonumber \\
&=& 1 + p_{u}^{eq}(e^{2\beta U_0} - 1) \nonumber \\
&=& 2(1 - p_{u}^{eq})  
\label{GJR_left1}
\end{eqnarray}
The right hand side of GJR is $\gamma = p_1 + p_2$, where $p_1$ and $p_2$ are 
probabilities to be determined by running the two protocols backwards for the case when the 
particle was observed in the up state in the forward cycle and in the down state in the
forward cycle respectively. Note that the time reversed protocols start with the system in the 
equilibrium state and no feedback is involved. $p_1$ is the probability of finding the particle
in the up state at time $t = \alpha$ with the state flipped at $t = \alpha$ (Note that the flip would in
general be carried out at $t = \alpha - \epsilon$, but we are looking at an engine with $\epsilon = 0$).
$p_1$ is thus the probability of finding the system in the down state in equilibrium, which is 
$(1 - p_{u}^{eq})$. $p_2$ is the probability of finding the particle 
in the down state at time $t = \alpha$, starting from equilibrium with no flip in the state. 
Thus $p_2$ is also given by $(1 - p_{u}^{eq})$. Thus the right hand side of GJR is given by,
\begin{equation}
    \gamma = 2 (1 - p_{u}^{eq}) \;.
    \label{GJR_right1}
\end{equation}
Comparing Eqs. (\ref{GJR_left1}) and (\ref{GJR_right1}) we see that GJR is valid for this
system.

Efficacy, $\gamma$, is a measure of how reversible the engine is. In the limit $\beta U_0 \gg 1$,
$p_{u}^{eq} \approx 0$ and $\gamma \approx 2$. This is the largest possible value for $\gamma$ 
for feedback process which involves two measurement outcomes. When $U_0 = 0$, $p_{u}^{eq} = \frac{1}{2}$ 
and efficacy becomes, $\gamma = 1$. For this case there is no feedback because the two states 
are identical and the flip does not make a difference to the state.
As expected, the GJR reduces to the usual Jarzynsky equality for this case. When $\beta U_0 \ll -1$,
efficacy becomes $0$. The information is used in least optimal manner in this situation. This is 
because the demon, rather than extracting work, flips the state when the system is in the lower of the 
energy states. Note that with $U_0 < 0$, the up state becomes the lower energy state.

\subsubsection{Verifying generalized integral fluctuation theorem}
To verify GIFT, we need the values of the information variable, $I$ and the unavailable information,
$I_u$ for the two outcomes of the measurement. For the case when the system is measured in the up state,
$I$ is given by $I_1 = -\ln{p_{u}^{eq}}$ and for the other case information is, $I_2 = -\ln(1- {p_{u}^{eq}})$. 
The unavailable information for the two cases are given by $I_{u1} = -\ln{p_1}$ and $I_{u2} = -\ln{p_2}$ 
respectively. Using the values of the probabilities determined above for the occurrence of the two outcomes, we have,
\begin{eqnarray}
\left <e^{-\beta W - I + I_u}\right> &=& e^{2\beta U_0 + \ln{p_{u}^{eq}} 
-\ln{(1-p_{u}^{eq})}} p_{u}^{eq} + (1 - p_{u}^{eq}) \nonumber \\
&=& 1 \;,
\label{GIFT_1}
\end{eqnarray}
and thus verifying GIFT for this case. Note that if one ignores the unavailable information, $I_u$, then GIFT
will be found to be violated. This is because we have assumed a measurement without error. In fact, one can 
easily see that $\left <e^{-\beta W - I}\right> = (1 - p_{u}^{eq})$, giving a value less than one.  

\subsection{Feedback engine with delay time ($\epsilon \ne 0$)}\label{sec2_2}
We now consider the case where there is a finite feedback delay time, $\epsilon$, between the
measurement and state flip. As discussed above, the relaxation time for the system is $\tau$. 
Delay in implementing the feedback would imply that at the instant of a state flip, there is a finite probability that 
the system's state differs from the measured state. These probabilities are:
\begin{eqnarray}
    P({\rm up}\;;\;t \;|\; {\rm up}\;;\;0) &=&  e^{-t/\tau}(1 - p_{u}^{eq}) + p_{u}^{eq} \nonumber \\
    P({\rm down}\;;\;t \;|\; {\rm up}\;;\;0) &=& 
    1-P({\rm up}\;;\;t \;|\; {\rm up}\;;\;0) \nonumber \\
    P({\rm down}\;;\;t\;|\; {\rm down}\;;\;0) &=&  e^{-t/\tau} p_{u}^{eq} + (1- p_{u}^{eq}) \nonumber \\
    P({\rm up}\;;\;t \;|\;{\rm down}\;;\;0) &=& 
    1-P({\rm down}\;;\;t \;|\; {\rm down}\;;\;0), \nonumber 
\end{eqnarray}
where we have defined $P(b\;;\; t_2 \;|\; a\;;\; t_1)$ as the probability that the state of the system is $b$
at time $t_2$ given that its state at time $t_1$ is $a$ ($t_2 > t_1$).

\subsubsection{Efficiency and work per cycle}
The average work extracted per cycle can be computed by taking into consideration the above probabilities.
The average work extracted per cycle is
\begin{equation}
   \displaystyle -\left<W\right> = 2 U_0 p_{u}^{eq} \tilde p - 2 U_0  p_{u}^{eq} (1 - \tilde p),
\end{equation}
where $\tilde p \equiv e^{-\epsilon/\tau}(1 - p_{u}^{eq}) + p_{u}^{eq}$. The first term in the above equation
accounts for the positive work extraction that happens when the state of the system is measured in the up state and
it is also in the up state at the time of the state flip. The second term corresponds to the negative work 
extracted, that happens when the state is measured to be up but has switched to down state during the delay 
time, $\epsilon$. 

In Fig. \ref{fig:U_pow} we have shown the variation of WPC with $U_0$ for different finite 
delay times: $\epsilon = 0.003$ (red dashed curve), $\epsilon = 0.010$ (green dotted curve), 
$\epsilon = 0.025$ (blue dash-dotted curve) and $\epsilon = 0.1$ (connected circles). The value 
of $\tau$ is taken to be $0.02$ for all the cases. As expected WPC is reduced for larger delay 
times because the information gained is utilised less optimally with increasing delay time. 
Also observed is the shift in the location of the peak value of WPC to smaller $U_0$ as delay 
time is increased. This means that for a fixed value of $U_0$, the maximum of WPC occurs for 
larger temperatures as delay time is increased. 
For large delay times compared to $\tau$, WPC becomes negative (connected circle curve for 
$\epsilon = 0.1$ in Fig. \ref{fig:U_pow}) indicating that most of the times when the state 
is flipped, it is in the down state. At intermediate delay times, the WPC takes both positive 
and negative values with the WPC values initially increasing from zero and then becoming 
negative and eventually approach zero from below (see dash-dotted curve for $\epsilon = 0.025$ in Fig. \ref{fig:U_pow}).

The efficiency of the information engine with feedback delay is given by,
\begin{equation}
\displaystyle \eta = \frac{2 \;U_0 \;p_{u}^{eq}\;(2 \tilde p - 1)}{k_B T \left<I\right>}.
\label{efficiency_1}
\end{equation}
The average information, $\left<I\right>$ is the same as that given in the previous section. 
This is because the feedback delay time has no bearing on the measurement probability 
when the cycle time is large compared to the relaxation time and the feedback delay time.
Fig. \ref{fig:U_eta} shows the variation of efficiency as a function of $U_0$ for the same
set of values for $\epsilon$ considered above for the case of WPC. Relaxation time, $\tau = 0.02$,
is also the same. We have seen that for $\epsilon = 0$, the efficiency increases monotonically with $U_0$, 
attaining the maximum value of $1$ as $U_0$ tends to infinity. But as the delay time is increased, the peak
in efficiency shifts to lower values of $U_0$. As expected, the peak value of efficiency also decreases 
as $\epsilon$ is increased. For $\epsilon$ of the order of $\tau$, both efficiency and WPC are maximum 
for $U_0 \lesssim 1$. These features are seen in the green dotted curve ($\epsilon = 0.010$) and 
the blue dash-dotted curve ($\epsilon = 0.025$) in Figs. \ref{fig:U_pow} and \ref{fig:U_eta}. 
%The peak is not visible in the dashed curve ($\epsilon = 0.003$) 
%as it is outside the range of $U_0$ displayed. 

\subsubsection{Generalized Jarzynski relation with feedback delay}
We have seen in the previous section that without feedback delay, the
efficacy, $\gamma = 2(1 - p_{u}^{eq})$. It was shown that this was indeed equal to
$\displaystyle \left<e^{-\beta W}\right>$ thus verifying GJR. We now find $\gamma$ for the
case with non-zero delay time and propose to verify the validity of GJR for this case.

For the present case, in the expression for $\gamma = p_1 + p_2$, $p_1$ is the probability
of finding the system in the up state at $t = \alpha$ with the system starting from equilibrium at $t=0$ and
a flip of the state being carried out at $t = \alpha - \epsilon$ (Note that there is no measurement involved
in the reverse process.). Thus $p_1$ is given by the sum of two terms: (i) the probability that at the 
time just before the flip, the system is in the down state (which means after the flip, the system 
will be in the up state) and then it remains in the up state till time $\alpha$ and (ii) the probability that
at the time just before the flip, the system is in the up state (which means after the flip, the system
will be in the down state) and then to be found in the up state at the time $\alpha$. $p_2$ on the other hand is
just the probability of the system to be found in the down state in equilibrium. This is the reverse
process when the particle in measured in the down state in the forward process and does not involve any
feedback. Thus we have,
\begin{eqnarray}
p_1 &=& (1 - p_{u}^{eq})\left[e^{-\epsilon/\tau}(1 - p_{u}^{eq}) + 
p_{u}^{eq}\right]  \nonumber \\ &&\hspace{0.1cm} + \; p_{u}^{eq}\left[p_{u}^{eq}
(1 - e^{-\epsilon/\tau})\right] \nonumber \\
p_2 &=& 1 - p_{u}^{eq} \;.
\label{p1witheps}
\end{eqnarray}
This gives,
\begin{equation}
    \displaystyle \gamma = 1 + e^{-\epsilon/\tau}(1 - 2p_{u}^{eq}),
    \label{gamma2}
\end{equation}
which reduces to $\gamma = 2(1 - p_{u}^{eq})$ for the case when $\epsilon = 0$, as expected.

To evaluate the LHS of GJR, note that the possible values of $e^{-\beta W}$ are $e^{2 \beta U_0}$,
$e^{-2 \beta U_0}$ and $1$, with probabilities $p_{u}^{eq} \tilde p$, $p_{u}^{eq} (1 - \tilde p)$ and
$(1 - p_{u}^{eq})$. Therefore,
\begin{equation}
\left <e^{-\beta W}\right> = 
e^{2\beta U_0} p_{u}^{eq} \tilde p + e^{-2 \beta U_0} p_{u}^{eq} (1 - \tilde p) + (1 - p_{u}^{eq}).
\end{equation}
Substituting for $\tilde p$ and making use of the relation $e^{2\beta U_0} = \frac{1 - 
p_{u}^{eq}}{p_{u}^{eq}}$, the above expression reduces to
\begin{equation}
 \left <e^{-\beta W}\right> = 1 + e^{-\epsilon/\tau}(1 - 2p_{u}^{eq})
    \label{GJR_2a}
\end{equation}
which is the same as $\gamma$ (Eq. \ref{gamma2}), thus validating the GJR. One can similarly verify the
validity of GIFT, which is presented in the appendix \ref{appendixA}.

\subsection{Comparison with simulation results}\label{sec2_3}
\begin{figure}
\includegraphics[scale=0.41]{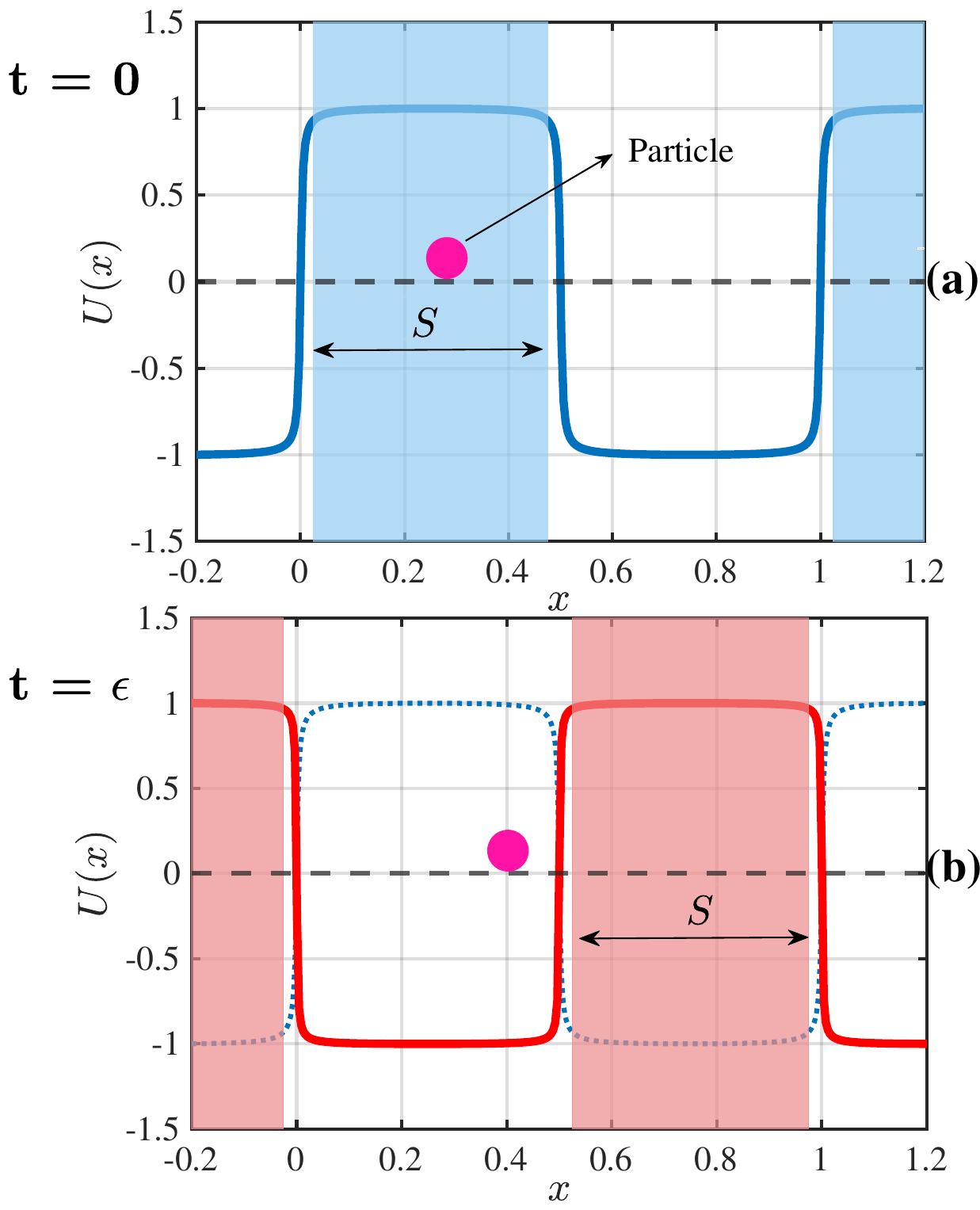}
\caption{\label{fig:potential} The working of the information engine using the motion of particle
in the square potential. When the particle is spotted in the blue regions of width $S$ at time 
$t = n \alpha$ (as in \textbf{(a)}), we flip the potential at $t = n \alpha + \epsilon$, as shown in \textbf{(b)}. 
The shaded areas in red indicate the new regions of potential maxima. The process is then repeated periodically.}
\end{figure}

Consider a particle moving in one dimension in a periodic square potential 
\begin{eqnarray}
   U_{s}(x) &=& \;\;U_0 \;\;\;(0 < x \le 0.5)  \nonumber \\
   &=& -U_0 \;\;(0.5 < x \le 1),
\end{eqnarray}
with $U_{s}(x) = U_{s}(x+1)$  as shown in Fig. \ref{fig:potential}. The particle is in contact with a
heat bath at temperature, $T$. One can implement an information engine using this system by
measuring the position of the particle and initiating a feedback protocol \cite{PhysRevE.106.054146}. 
The protocol closely resembles that of the two state information engine discussed above and is as follows: 
At times given by $t = n \alpha$, a measurement of the particle’s position is carried out. If the particle 
is located in the region with higher potential energy, referred to as region $S$ (see Fig. \ref{fig:potential} (a)), 
then the potential is flipped (that is, $U_{s}(x) \rightarrow -U_{s}(x)$) at a time 
$t = n \alpha +  \epsilon$ where $\alpha$ is the engine cycle time and $\epsilon$ 
is the feedback delay time. If the particle is not spotted in $S$, 
then no feedback process is initiated. The interval $\alpha - \epsilon$ is kept large enough to 
ensure that the system equilibriates before each measurement. This is not a necessary part of the 
current model but is done so that the comparison with the analytical results from the two sate model can be made.

Even though the state space of the current system, which is a continuum of states, 
is different from that of the two level system considered above, there are similarities. 
In equilibrium, the probability for finding the particle in the region of higher 
potential will be equal to the probability of finding the two level system in 
the up state. In the two-state model, the relaxation to equilibrium is governed 
by a single relaxation time. However, this process might differ for the particle in the 
square potential and could involve multiple time scales whose values depend on the height 
and the period of the potential. But as a first approximation, one can model the relaxation 
using a single relaxation time approximation. This would allow us to 
compare the simulation results for the particle in the square potential with the analytical 
results for the two level system by using the relaxation time as a fit parameter. 
We carry out this comparison below.

The simulation has been carried out using the over-damped Langevin equation,
\begin{equation}
\dot{x} = \frac{F(x)}{m\xi} + \frac{\zeta}{m\xi} \;,
\label{bdeq}
\end{equation}
where $m$ is the mass of the particle and $\xi$ is the friction coefficient. 
$\zeta(t)$ is the thermal noise with zero average and the correlation function is given by 
$\left<\zeta(t) \zeta(t')\right> = \Gamma \delta(t - t')$. Fluctuation-dissipation relation
connects the strength of the noise, $\Gamma$, to the friction coefficient, $\xi$, by the 
relation: $\Gamma = 2 m \xi k_{B}T$. $F(x)$ is the conservative force arising from a potential, $U(x)$.
The square potential is modelled using the function 
$\displaystyle U(x) = C(\Delta) \tan^{-1}\left[{\frac{\sin{(2\pi x)}}{\Delta}}\right]$.  
The value of the parameter $\Delta$ determines the sharpness of the potential and the parameter 
$C$ is adjusted so as to make the amplitude of the potential to be $U_0$. Fig. \ref{fig:potential}
shows the shape of $U(x)$, for $\Delta = 0.02$ and $C = 0.64$ which gives a good approximation to
$U_s(x)$ with $U_0 = 1$.

The above equation Eq. (\ref{bdeq}) has been integrated numerically 
using the discretized version \cite{ermak},
\begin{equation}
x(t + \delta t) = x(t) + \frac{F(x)}{m\xi} \delta t +  f_g, 
\label{ermak}
\end{equation}
where $\delta t$ is the time step and $f_g$ is a Gaussian distributed random variable with 
zero mean and variance equal to $\frac{2 k_B T}{m\xi} \delta t$. Since $U(x)$ does not have
a discontinuity at $x = 0.5$, one needs to choose the region $S$ appropriately.
We have chosen region $S$ such that it approximately covers the elevated part of the potential 
(see Fig. \ref{fig:potential}). $S$ is taken as the region between $x = 0.025$ and
to $x = 0.475$ before the flip (Fig. \ref{fig:potential} (a)) and from $x = 0.525$ to 
$x = 0.975$ after the flip (Fig. \ref{fig:potential} (b)), encompassing a total length of $0.45$ 
and periodically repeating. We work with a system of units defined by $\xi = 1$, $m = 1$ 
and $k_{B}T = 1$. Time scale in the problem is set by 
$\xi^{-1}$, which is set to $1$. The length scale of the problem is the period of the potential,
which is $1$. The integration time step of the simulation is taken to be $\delta t = 10^{-5}$. The time step
has been kept small because at the region where the potential changes, close to $mod(x,1) = 0.5$,
the forces can be very large. We have verified the convergence of the solution by checking
out sample trajectories at one order smaller time step. Simulations have been carried out 
by varying the amplitude of the potential $U_0$ as well as the
delay time $\epsilon$. The WPC, efficiency and efficacy are computed by averaging over 
$10^6$ cycles. The averaging over large number of trials are particularly necessary for 
finding efficacy accurately \cite{jarzynski2011equalities,liphardt2002equilibrium}.

\begin{figure}
\includegraphics[scale=0.45]{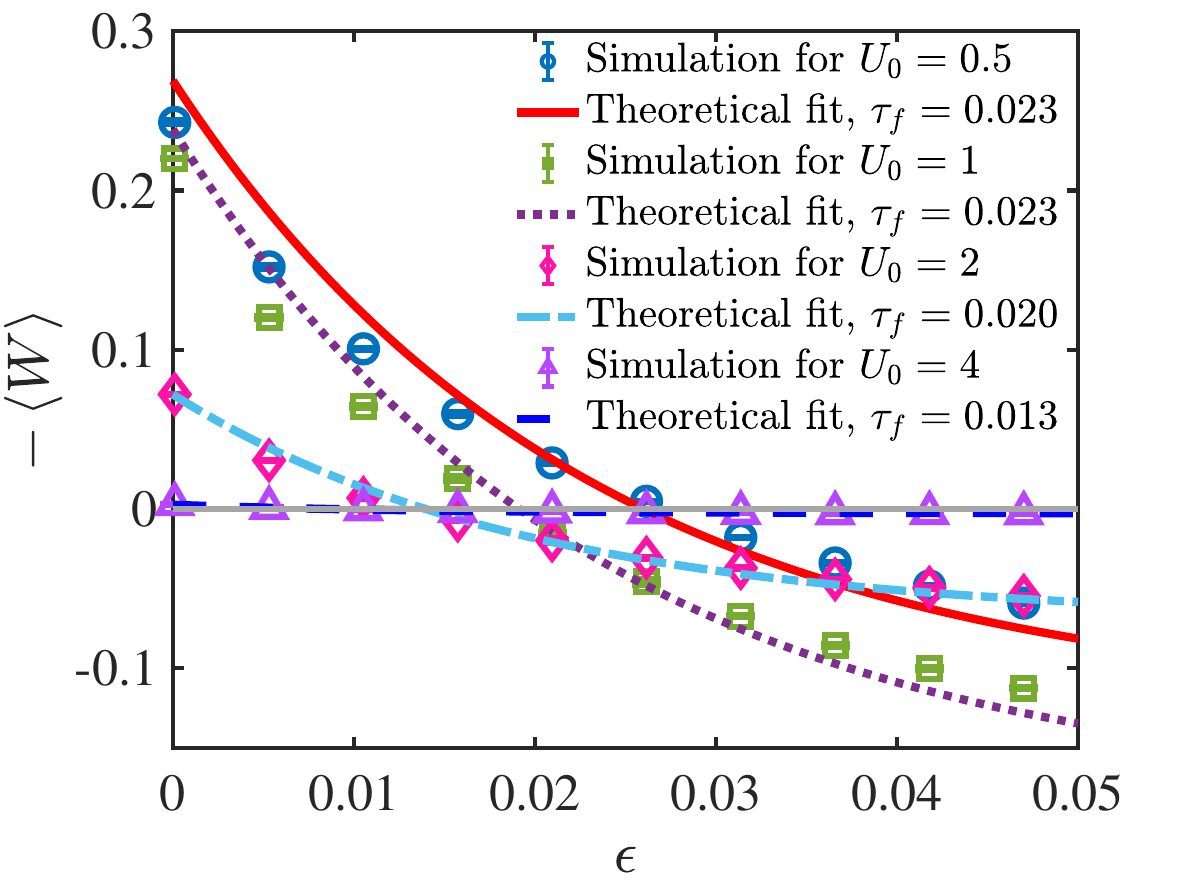}
\caption{\label{fig:eps_pow} Variation of WPC as a function of delay time $\epsilon$. The data
points are from the simulation of the information engine based on particle moving in the square potential.
Different symbols correspond to various values of the amplitude, $U_0$. The continuous curves are the fits
using the results derived for the two state model with $\tau$ as the fitting parameter. 
The error bars in the figure are standard deviations.}
\end{figure}
 
\begin{figure}
\includegraphics[scale=0.45]{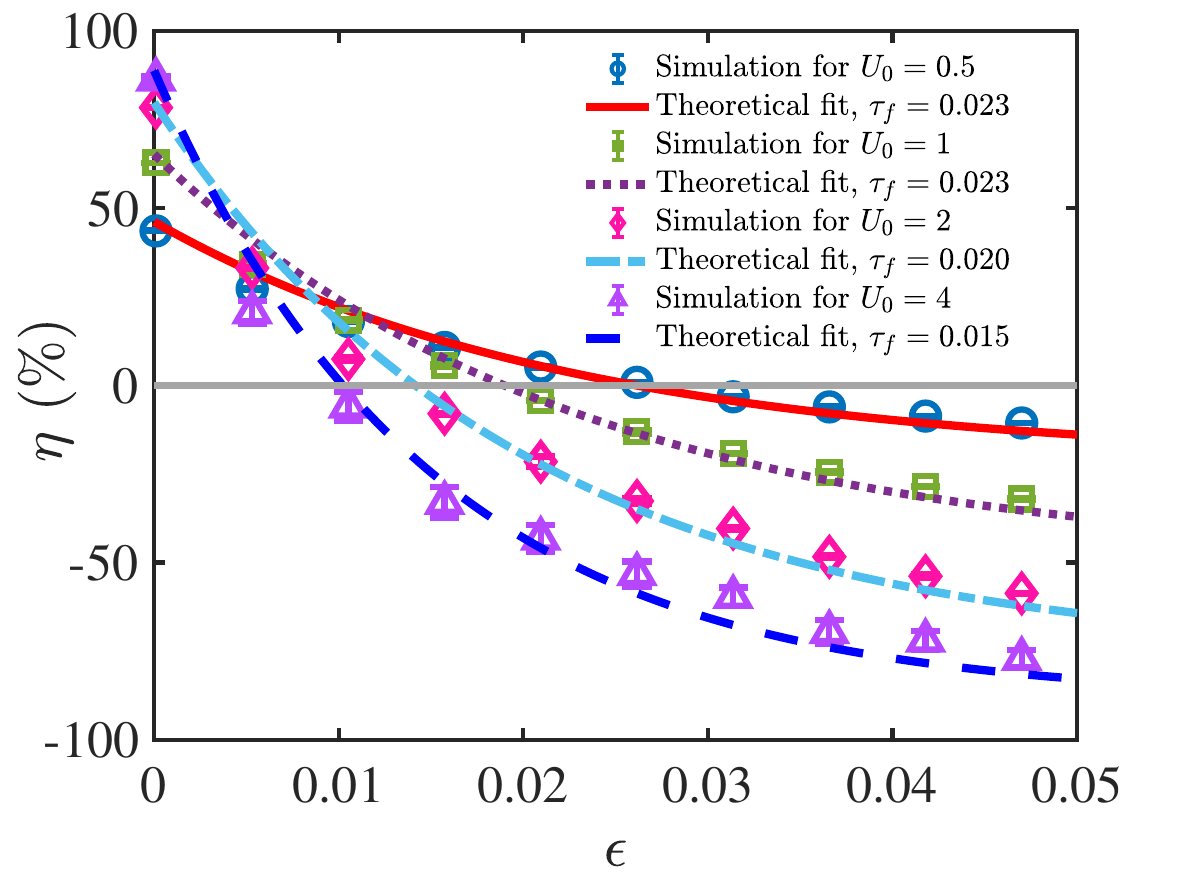}
\caption{\label{fig:eps_eta} Variation of efficiency as a function of delay time $\epsilon$. The data
points are from the simulation of the information engine based on particle moving in the square potential.
Different symbols correspond to different values of the amplitude, $U_0$. The continuous curves are the fits
using the results derived for the two state model with $\tau$ as the fitting parameter. 
The error bars in the figure are standard deviations.}
\end{figure}

Work per cycle as a function of the delay time for various values of $U_0$ are shown in 
Fig. \ref{fig:eps_pow}. As expected, for a given value of $U_0$, WPC decreases
with $\epsilon$ because the particle might drift away from the higher potential region if
one waits longer for the feedback process after the measurement. It is observed that the zero
crossing of WPC occurs at lower values of $\epsilon$ for larger $U_0$ values. This implies 
that the range of delay time over which the information engine can extract work decreases 
with decreasing temperature. For small $U_0$, the WPC reduces by a factor of almost $0.5$ when
$\epsilon$ is varied from $0$ to $0.01$, which is approximately half the relaxation time
(see red solid line in Fig. \ref{fig:eps_pow}). This drop is more drastic for higher values of 
$U_0$ with WPC dropping by nearly one fourth its value at $\epsilon = 0$
(see blue dash-dot line in Fig. \ref{fig:eps_pow}). The curves in Fig. \ref{fig:eps_pow} are theoretical fit to the 
data obtained using the analytic results from the two state model. The relaxation time, 
$\tau$, is the fit parameter. We see that there is good fit for all values of $U_0$ 
considered. This justifies the single relaxation time approximation. %The value of $\tau$ 
%for different values of $U_0$ is different with $\tau$ decreasing with increasing $U_0$. 
It is seen that value of $\tau$ depends on $U_0$ with $\tau$ decreasing as $U_0$ increases.
The dependency of efficiency on $\epsilon$
is given in Fig. \ref{fig:eps_eta} for different values of $U_0$. Like in the case of WPC the efficiency
decreases monotonically with delay time. The theoretical fits using the two state results yield similar values of $\tau$ 
as those obtained from the WPC curve fits.

\begin{figure}
\includegraphics[scale=0.45]{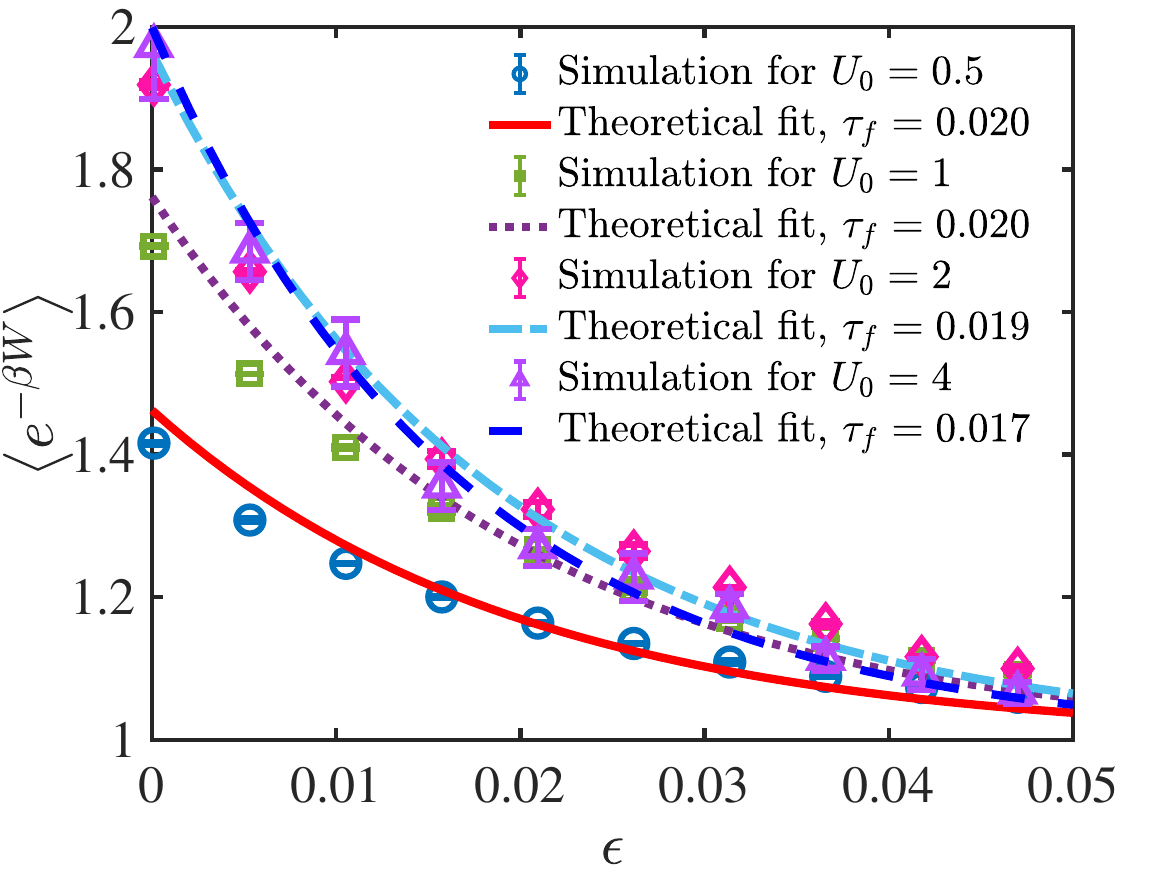}
\caption{\label{fig:fit} Variation of efficacy as a function of delay time $\epsilon$. The data
points are from the simulation of the information engine based on particle moving in the square well potential.
Different symbols correspond to different values of the amplitude, $U_0$. The continuous curves are the fits
using the results derived for the two state model with $\tau$ as the fitting parameter. 
The error bars in the figure are standard deviations. The maximum value that efficacy can have is $2$ and is
attained for short delay times and large $U_0$.}
\end{figure}

Efficacy, $\gamma$, has been computed by finding the average $\left<e^{-\beta W)}\right>$ from the simulations.
The dependency of $\gamma$ on the delay time for various values of $U_0$ are shown in Fig. \ref{fig:fit}.
The plot qualitatively resembles those found in the experimentally realized information engine based on a particle 
moving in a sinusoidal potential \cite{toyabe2010experimental}. The maximum value that efficacy can take is $2$
for this feedback engine as there are two outcomes possible during the measurement. High efficacy values are
obtained for small delay times and large $U_0$. For $\epsilon \approx 0$ and $U_0 \geq 4$, we find efficacy
values close to $2$. As expected, the efficacy approaches one as the delay time is increased implying that the
Jarzynski equality holds when the feedback is redundant. The theoretical fit for this case leads to values 
of $\tau$ close to those obtained from the previous fits. %The GIFT for the same system is confirmed by calculating 
%the average $\displaystyle \left <e^{-\beta W - I + I_u}\right>$ through simulations, as detailed in the appendix \ref{appendixB}.
%\cite{}

\begin{figure}
\includegraphics[scale=0.45]{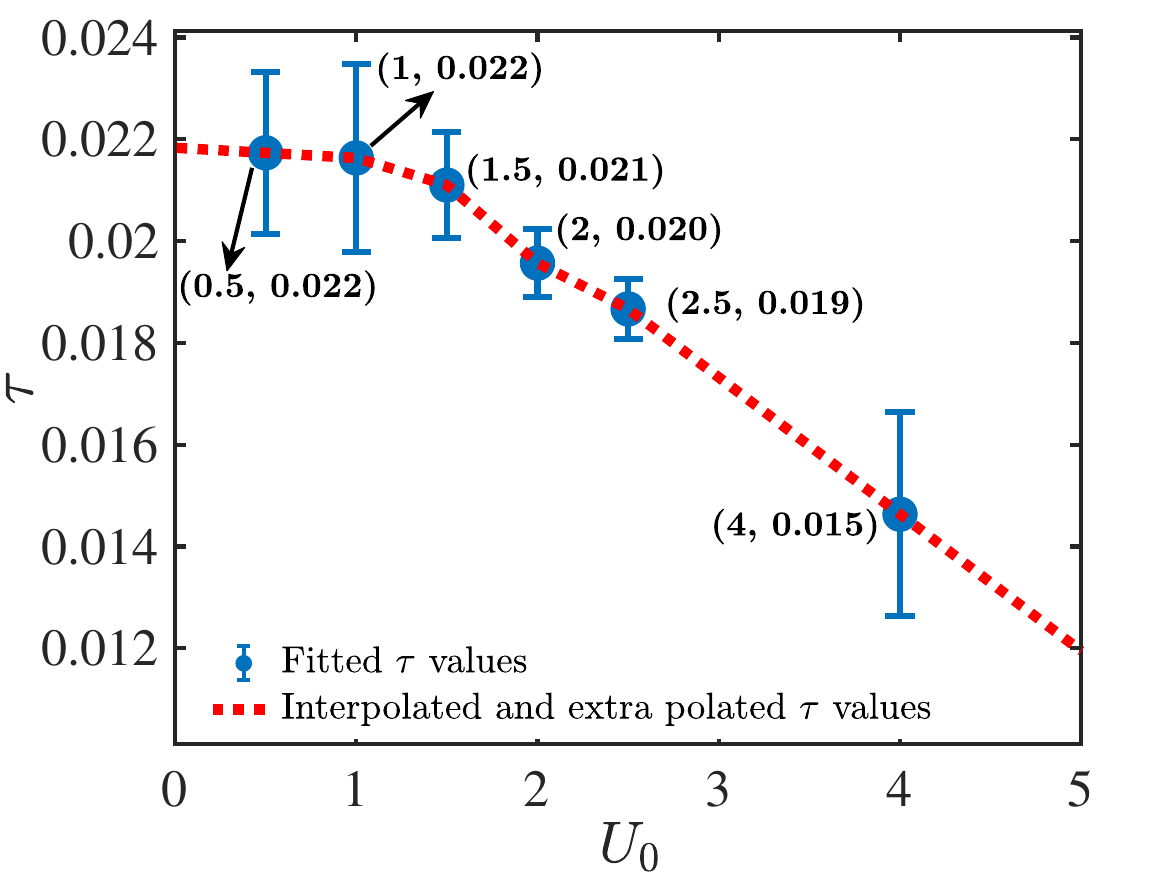}
\caption{\label{fig:tau_extrapol} The variation of the fitted value of relaxation time as a function
of $U_0$ (in units of $k_B T$). Each data point is the average of $\tau$ obtained from fitting the WPC, efficiency and
efficacy as a function of $\epsilon$ given in Figs. \ref{fig:eps_pow}, \ref{fig:eps_eta}, and \ref{fig:fit}. 
The red dotted line is the value of $\tau$ for values of $U_0$, extrapolated from the available data.
The error bars in the figure are standard deviations.}
\end{figure}

\begin{figure}
\includegraphics[scale=0.45]{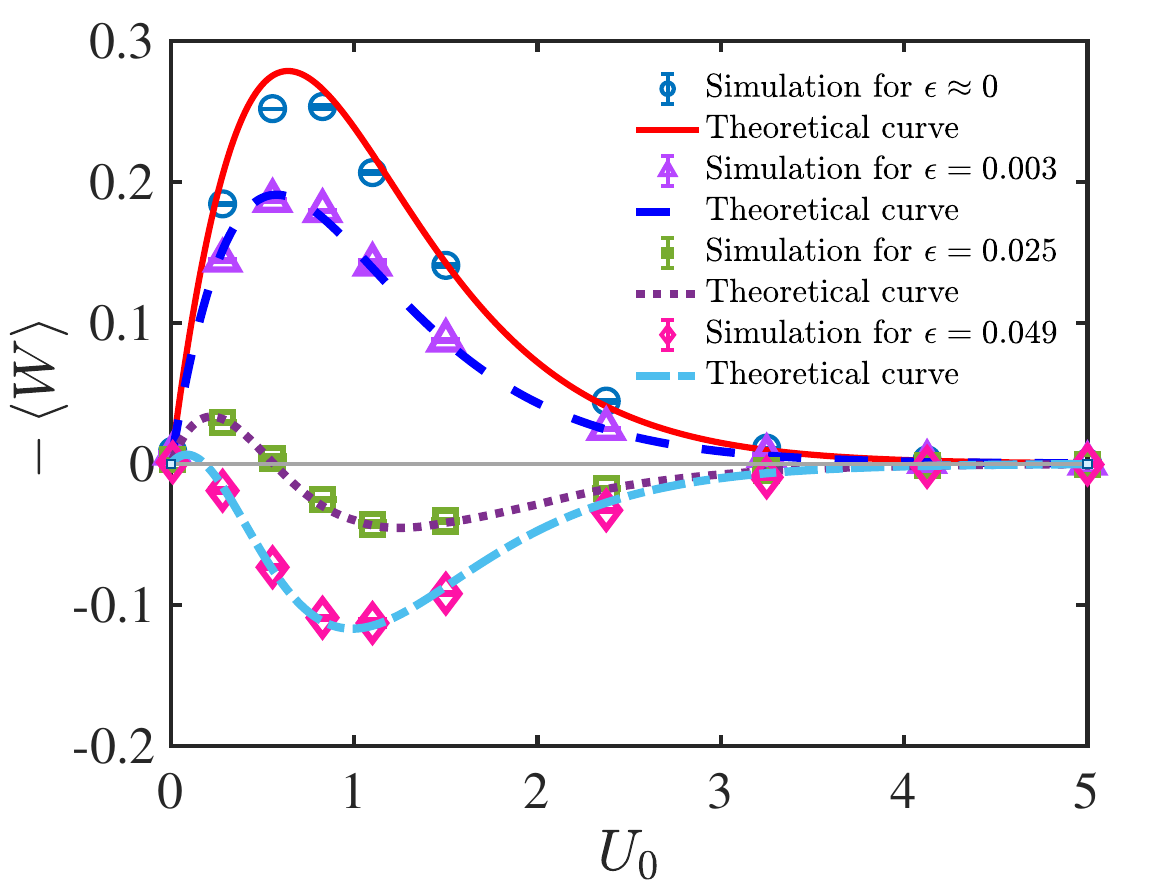}
\caption{\label{fig:U_pow_comp} Work done per cycle as a function of amplitude of the square potential
(in units of $k_B T$) for the information engine based on the particle moving in the square potential. 
The data points are from the simulation results. The lines are plotted using the analytical results 
from the two state model. The relaxation times for various value of $U_0$ are obtained from the 
extrapolation curve in Fig. \ref{fig:tau_extrapol}. The error bars in the figure are standard deviations.}
\end{figure}

\begin{figure}
\includegraphics[scale=0.45]{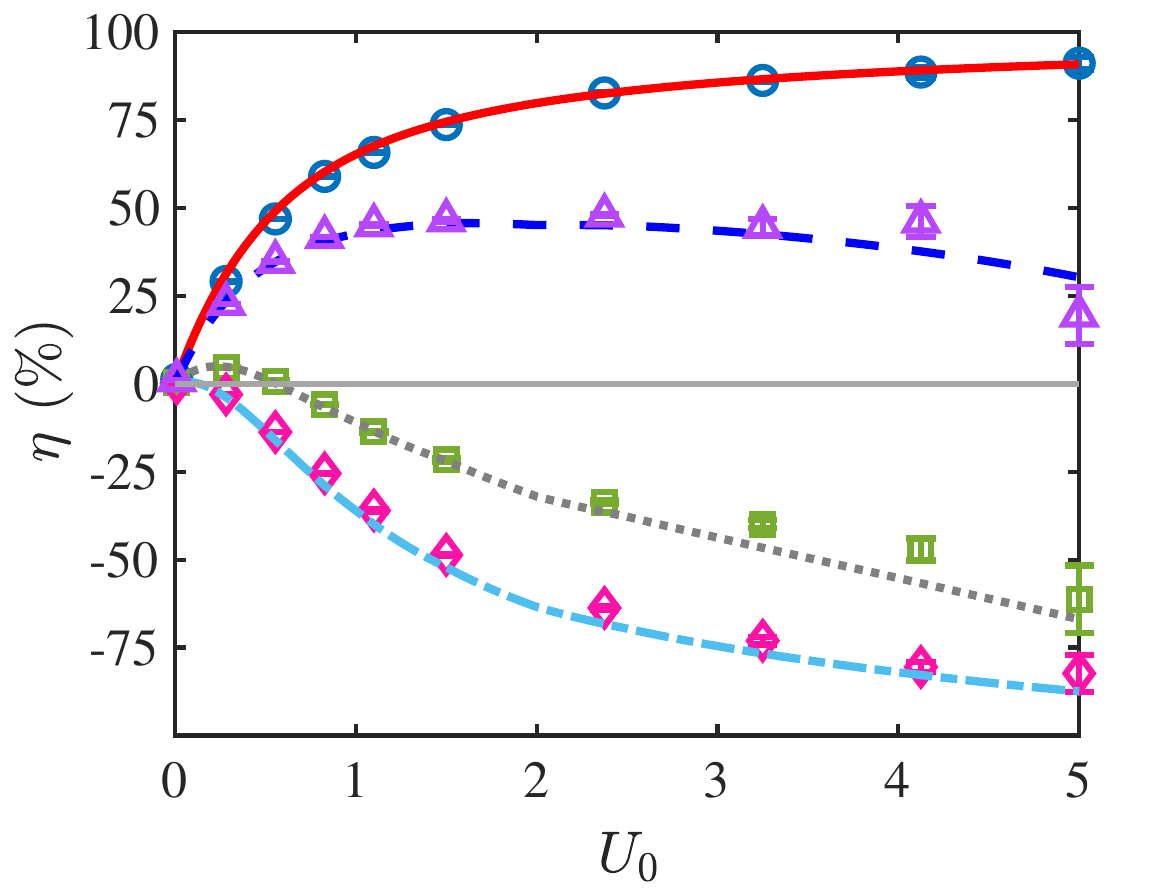}
\caption{\label{fig:U_effic_comp} Efficiency as a function of amplitude of the square potential (in units of $k_B T$)
for the information engine based on the particle moving in the square potential. The data points are 
from the simulation results. The lines are plotted using the analytical results from the two state model. 
The relaxation times for various value of $U_0$ are obtained from the extrapolation curve 
in Fig. \ref{fig:tau_extrapol}. The error bars in the figure are standard deviations. 
The legends used are the same as that in Fig. \ref{fig:U_pow_comp}.}
\end{figure}

We have averaged the values of $\tau$ obtained from the three sets of fittings carried out above.
This gives the variation of $\tau$ as a function of $U_0$ and is shown in Fig. \ref{fig:tau_extrapol}.
Further, we extrapolate this data to get estimates of $\tau$ for arbitrary values of $U_0$ close
to the ones used in the simulations. It is seen that the relaxation time has a strong dependence on 
the amplitude of the potential, particularly for large values of $U_0$. We have used the
extrapolated data of $\tau$ values to plot the theoretical curves for WPC vs. $U_0$ and $\eta$ vs. $U_0$  along with
the results obtained from simulation. These results are shown in Figs. \ref{fig:U_pow_comp} and 
\ref{fig:U_effic_comp} respectively. There is excellent match between simulation data and model results.
Note that the plots of WPC vs. $U_0$ and $\eta$ vs. $U_0$ in Figs. \ref{fig:U_pow} and 
\ref{fig:U_eta} respectively are for fixed $\tau$ value, whereas for the present case $\tau$
varies with $U_0$. It is seen from $\eta$ vs $U_0$ plot (Fig. \ref{fig:U_effic_comp}) that for short delay 
times compared to the relaxation time the efficiency increases with $U_0$ (red solid curve). But at larger
delay times the efficiency is maximum at intermediate values of $U_0$ (blue dashed curve and black dotted curve).


\section{Conclusion}\label{sec3}
Information engine based on a two state model is possibly the simplest information
engine that can be studied with feedback delay time incorporated. The model allows for
an exact calculation of important engine performance indicators: efficiency, work extracted per cycle
and efficacy. The key control parameters of the engine are the feedback delay time and the energy gap 
between the levels or alternately the temperature at which the engine functions. Some of the important
observations from the analytical study of the two state system with feedback are the following:\\
{\bf (i)} The engine performance deteriorates with increasing delay time. As delay time becomes large 
compared to the relaxation time, one would expect the work extracted \\
per cycle to saturate to a negative value. This is because the system is more likely to be in
the down state ($(1 - p_{u}^{eq}) > p_{u}^{eq}$) at the time of the state flip. The efficacy in this
limit will saturate to value $1$, indicating that the information gained is not utilized in
extracting work from the thermal fluctuations.\\
{\bf (ii)} For the case of zero delay time the efficiency increases monotonically with $U_0$ whereas
WPC has a peak at intermediate $U_0$. For finite delay time the peak in WPC and $\eta$ shift to lower
values of $U_0$. \\
{\bf (iii)} The range of delay time over which the engine can extract positive work increase as $U_0$
decreases, or equivalently the range increases with temperature. \\
{\bf (iv)} For delay time of the order of the relaxation time, both WPC and $\eta$ have maxima close to
but below $U_0 = 1$ (in units of $k_B T$). Since $2 U_0$ is the level spacing, this implies that for a 
fixed level spacing the optimal temperature to run the engine is roughly given by $k_B T$ is of the same
order as the energy difference between the levels.

The information engine presented here allowed for verification of generalized fluctuation theorems 
of stochastic thermodynamics. The importance of the unavailable information term in the GIFT relation 
is explicitly brought out. It is to be noted that the introduction of error into the measurement 
process will alter the form of the GIFT. In that scenario, the $I_u$ in the LHS of the GIFT will not be present. 
In the context of stochastic thermodynamics of feedback systems, this simple model can also be of pedagogic 
interest to understand various fluctuation theorems.

Comparison of the model results with simulation of an information engine based on a particle 
moving in square potential leads to good match. The efficacy variation as a function of delay 
time shows the same features that were observed in the experimentally realized information engine based
on a particle moving in a sinusoidal potential \cite{toyabe2010experimental}. The fit values of
the relaxation time determined from variation of efficiency, WPC and efficacy with $\epsilon$, all give
similar values of $\tau$ for all values of $U_0$ considered. The variation of WPC 
and efficiency with $U_0$ for the particle based information engine has similar features 
as that for the two state system. The details of the behavior
are however different due to the dependence of relaxation time on the amplitude of the
square potential.

The current work assumes that the cycle time is large compared to the relaxation time, so that one
can assume equilibrium conditions at the beginning of each cycle. This is the reason one had
to look at WPC, rather than power of the engine. One can extend the analysis to the case where
cycle time is finite. One then needs to work out the steady state probability with feedback in place.
Another improvement to the model could be introduction of error in the measurement process. This
would make the model more realistic and will allow one to optimize the engine in the presence of
imperfect measurements. Work is in progress to incorporate these modifications to the model. 
Most of the results discussed here should be experimentally accessible in the framework of 
colloidal particle based information engines.

\acknowledgments
TJ would like to acknowledge financial support under the DST-SERB 
Grant No: CRG/2020/003646.

\appendix
\section{Generalized integral fluctuation theorem (GIFT) with feedback delay $\epsilon \neq 0$}\label{appendixA}

To show that GIFT holds, we need to prove $\left <e^{-\beta W - I + I_u}\right> = 1$. 
The possible values of $W$ are $-2U_0$, $2U_0$ and $0$, with probabilities $p_{u}^{eq} \tilde p$, 
$p_{u}^{eq} (1 - \tilde p)$ and $(1 - p_{u}^{eq})$ respectively. The corresponding values of
$I$ are  $-\ln{p_{u}^{eq}}$, $-\ln{p_{u}^{eq}}$ and $-\ln(1-{p_{u}^{eq}})$ respectively. And the values
for $I_u$ are $-\ln{p_{1}}$, $-\ln{p_{1}}$ and $-\ln{p_{2}}$ respectively (where $p_1$ and $p_2$ are
given by Eq. (\ref{p1witheps})).
\begin{eqnarray}
\displaystyle \left <e^{-\beta W - I + I_u}\right> &=& e^{2\beta U_0 + \ln{p_{u}^{eq}}-\ln{p_1}}p_{u}^{eq}\tilde p \nonumber \\ &&\hspace{0.1cm}
+ e^{-2\beta U_0 + \ln{p_{u}^{eq}}-\ln{p_1}}p_{u}^{eq}(1-\tilde p) \nonumber \\ &&\hspace{0.1cm}
+ e^{0+\ln{(1-p_{u}^{eq})-\ln{p_2}}} (1 - p_{u}^{eq}) 
\end{eqnarray}
Substituting for $p_2 = 1 - p_{u}^{eq}$ and simplifying,
\begin{eqnarray}
\displaystyle \left <e^{-\beta W - I + I_u}\right> &=&  e^{2\beta U_0}\frac{{p_{u}^{eq}}^2}{p_1} \tilde p + e^{-2\beta U_0}\frac{{p_{u}^{eq}}^2}{p_1} (1-\tilde p) \nonumber \\ &&\hspace{0.1cm}
+ 1 - p_{u}^{eq}.
\end{eqnarray}
Substituting for $\tilde p$ and $p_1$ and making use of the relation $e^{2\beta U_0} = \frac{1 - 
p_{u}^{eq}}{p_{u}^{eq}}$ and $e^{-2\beta U_0} = \frac{p_{u}^{eq}}{1 - 
p_{u}^{eq}}$, the above expression reduces to
\begin{eqnarray}
    \displaystyle \left <e^{-\beta W - I + I_u}\right> &=& c_1\left[p_{u}^{eq}+e^{-\epsilon/\tau}(1-2p_{u}^{eq})\right]  \nonumber \\ 
    &&\hspace{0.1cm} + 1-p_{u}^{eq}, \;
    \label{gift_eps}
\end{eqnarray}
where $\displaystyle c_1 = \frac{p_{u}^{eq}}{p_{u}^{eq}+e^{-\epsilon/\tau}(1-2p_{u}^{eq})}$.\\\\
Simplifying \ref{gift_eps} we get,
\begin{eqnarray}
    \displaystyle \left <e^{-\beta W - I + I_u}\right> = 1
\end{eqnarray}

\bibliographystyle{apsrev4-2}
\bibliography{References}%apssamp Produces the bibliography via BibTeX.

\end{document}